\documentclass[prb,preprint,amsmath,showpacs]{revtex4}
\usepackage{graphicx}

\begin{document}

\title{Single crystal growth and investigation of $\bf Na_xCoO_2$ and $\bf Na_xCoO_2~yH_2O$}
\author{D. P. Chen, H. C. Chen, A. Maljuk, A. Kulakov, H. Zhang, P. Lemmens, and C. T. Lin*}
\address{Max Planck Institute for Solid State Research, Heisenbergstr. 1,
D-70569 Stuttgart, F. R. Germany }

\footnote{*Corresponding author, Phone: 0049-711-689 - 1093, email:
ct.lin@fkf.mpg.de}

\pacs{(Keywords: $\rm Na_xCoO_2 yH_2O$ single crystals, floating zone
growth, intercalation and hydration, superconductivity, transport
properties)}

\begin{abstract}
A systematic study of $\rm Na_xCoO_2$ (x=0.50 - 0.90) and $\rm Na_xCoO_2
yH(D)_2O$ (\emph{x}=0.26 - 0.42, \emph{y}=1.3) has been performed to
determine phase stability and the effect of hydration on the structural and
superconducting properties of this system. We show that a careful control
of the Na deintercalation process and hydration dynamics is possible in
single crystals of this system. Furthermore, we give experimental evidence
that the dependence of the superconducting transition temperature on Na
content is much weaker than reported earlier. Implications of this effect
for the understanding of the superconducting phase diagram are discussed.


\end{abstract}

\maketitle

\section{Introduction}
The recently discovered superconductor $\rm Na_xCoO_2 yH_2O$ has attracted
considerable attention as being the first superconducting cobaltite and due
to evidence for important electronic correlations \cite{1}. Its
superconducting transition temperature of maximum $\rm T_c\approx$5~K
exhibits a composition dependence, decreasing for both underdoped and
overdoped materials, as observed in the cuprates. This suggests that a
detailed characterization of the electronic and magnetic behavior of this
new type of materials and their interplay with structural peculiarities may
contribute to a more fundamental understanding of the high $\rm T_c$
superconductivity in cuprates, such as the layered crystal structure
\cite{2}, magnetic susceptibility \cite{3,4}, electronic anisotropy
\cite{5}, irreversibility lines and superconducting parameters $\rm
\lambda_{ab}$, $\rm H_{c2}$.

The physical properties are determined by the amount of sodium and water in
$\rm Na_xCoO_2 yH_2O$ via their influence on the Co valence state and the
resulting local distortions of the $\rm CoO_6$ octahedra present in the
structure. The compound $\rm Na_xCoO_2$ shows manifestations of frustration
in its physical properties, because Co occupies a triangular lattice and
the exchange interactions are antiferromagnetic \cite{6,7}. The close
relationship between structural and electronic properties establishes
itself also on the Fermi-Surface that is expected to show features of
strong nesting \cite{tanaka03,johannes04}. Furthermore, there exists
evidence that superconductivity might have an unconventional order
parameter \cite{tanaka03}. When water is introduced into the compound, it
exhibits chemical and structural instabilities. The $\rm Na^{+}$ ions are
at least partially mobile at elevated temperatures and may order at low
temperatures in dependence of composition and sample treatment. The same is
proposed for the $\rm Co^{4+}/Co^{3+}$ charge state on the triangular
lattice \cite{huang04}.

All of these features make the study of the physical and chemical
properties of the compounds difficult. However, nearly all the studies
reported so far have been made on either sintered specimens or poorly
characterized polycrystalline samples. These materials are often found to
be inhomogeneous, especially with respect to the distribution of $\rm Na^+$
ions in the lattice and in the intergranular spaces, the $\rm Na_2O$
decomposition of the compound, and with respect to the intercalation of
water. In polycrystalline samples water may in part accumulate in
intergranular spaces. The possible presence of second phases leaves a
degree of uncertainty when interpreting the structure and the electrical
and superconducting properties as function of composition, i.e.
establishing a superconducting phase diagram. This problem may be
circumvented by the use of high quality single crystals whose chemical
composition and crystal structure can be properly determined.

When growing $\rm Na_xCoO_2$ single crystals, considerable difficulties
appear on account of the high $\rm Na_2O$ vapor pressure, which increases
exponentially from $\rm 10^{-5}$ to $\rm 10^{-3}$ torr with heating to
temperatures between 500 and 800 $\rm ^\circ$C, followed by a noticeable
evaporation. Therefore ceramic powders were synthesized using additional
$\rm Na_2CoO_3$  for compensation of the Na loss during heating  \cite{2,8}
or a "rapid heat-up" technique \cite{9} to avoid the formation of a
non-stoichiometric compound. Solution growth by using NaCl flux \cite{10}
was performed that unfortunately lead only to thin crystals ($<$0.03 mm) or
non-stoichiometric and possibly contaminated samples. To obtain the
superconducting phase the crystal must be hydrated by chemical oxidation
\cite{1}. The compound can be partially decomposed during the oxidation
process, leading to highly defective crystals containing Na-poor phases.

In this contribution we present a single crystal growth method to prepare
large and high quality single crystals and demonstrate their chemical,
thermal, and structural behavior and electric properties under dry and
humid conditions.

\section{Experimental}
Single crystals were grown in an optical floating zone furnace (Crystal
System Incorporation, Japan) with 4 x 300~W halogen lamps installed as
infrared radiation sources. Starting feed and seed materials were prepared
from $\rm Na_2CO_3$ and $\rm Co_3O_4$ of 99.9\% purity with the nominal
composition of $\rm Na_xCoO_2$, where \emph{x}=0.50, 0.60, 0.70, 0.75,
0.80, 0.85 and 0.90, respectively. Well-mixed powders were loaded into
alumina crucibles and heated at 750~$\rm ^\circ$C for a day. The heated
powders were reground and calcined at 850~$\rm ^\circ$C for another day.
They were then shaped into cylindrical bars of $\approx$ 6x100 mm by
pressing at an isostatic pressure of $\approx$70~Mpa and then sintered at
850 $\rm ^\circ$C for one day in flowing oxygen to form feed rods.

The sintered feed rod was premelted with a mirror scanning velocity of 27
mm/h by travelling the upper and lower shafts, respectively, to densify the
feed rod. After premelting the $\approx$20~mm rod long was cut and used as
a first seed and hereafter the grown crystal was used as a seed. The feed
rod and the growing crystal were rotated at 15~rpm in opposite directions.
In an attempt to reduce the volatilization of Na and obtain stoichiometric
and large crystals, we applied travelling rates of 2~mm/h under pure oxygen
flow of 200 ml/min throughout the growing procedure.

To obtain superconductivity the Na was partly extracted by placing crystals
in a 6.6~mol $\rm Br_2/CH_3CN$ solution during 100 hours and then washed
out the solution by $\rm H_2O$ or $\rm D_2O$ ~\cite{1}. Alternatively, the
electrochemical technique was also applied, using an aqueous solution of
NaOH as an electrolyte to partially extract Na \cite{11}. This technique
needs a longer time for deintercalation and the resulting superconducting
transitions are generally sharper.

Thermogravimetric and differential thermal analysis (TG-DTA) was performed
to study the melting behavior as well as the time and temperature
dependence of the water loss in the compound. Single crystal XRD
measurements were carried out with the X-ray diffractometer (Philips PW
1710) using Cu $K_{\alpha}$ radiation, a scanning rate of 0.02 degrees per
minute, and $\theta$ - 2 $\theta$ scans from 5 to 90 degrees. The lattice
parameters obtained from XRD data were refined using the commercial program
PowderCell. The chemical composition was determined by energy dispersive
X-ray (EDX) including microanalysis. The as-grown crystals were cleaved
parallel to the \emph{a} axis and the composition of Na and Co was
determined across the cleaved section.

\begin{figure}[tbp]
\includegraphics[width=0.7\linewidth]{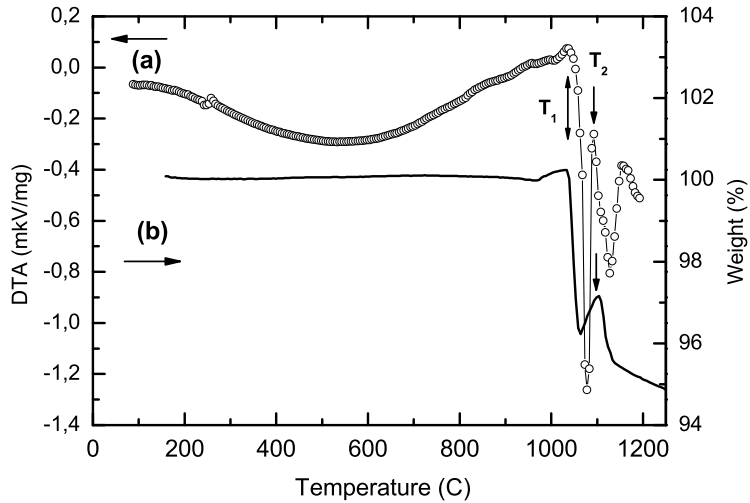}
\caption{ DTA-TG analysis by melting single crystalline $\rm Na_{0.7}CoO_2$
heated at 7.5 $\rm ^\circ$C/min up to 1200-1250 $\rm ^\circ$C in flowing
oxygen. $\rm  T_{1}$=1035 $\rm ^\circ$C, $\rm T_{2}$=1092 $\rm ^\circ$C.
(a) The melting behavior of the compound, (b) temperature dependence of the
weight loss.} \label{1}
\end{figure}

\section{Results and discussions}
\subsection{DTA-TG analysis}
The melting behavior of a $\rm Na_{0.7}CoO_2$ crystal was
investigated by DTA-TG measurements and with a high temperature optical
microscope. A small single crystal was loaded in a Pt crucible and then
heated in the DTA-TG apparatus (NETZSCH STA-449C) at 7.5 $\rm ^\circ$C/min
up to 1200-1250 $\rm ^\circ$C in flowing oxygen. Two peaks with onset $\rm
T_1$=1035 $\rm ^\circ$C and $\rm T_2$=1092 $\rm ^\circ$C on the DTA curve
correlate very well with a weight loss observed by TG, as shown in Figs.
\ref{1}(a) and (b), respectively. No weight changes were observed below 980
$\rm ^\circ$C.

The high temperature optical microscope study of the crystal reveals that
the liquid phase appears at $\rm T_1$=1035 $\rm ^\circ$C but the solid
phase (crystal) still remains up to $\rm T_2$=1092 $\rm ^\circ$C,
exhibiting a coexistence of solid and liquid phases. Our investigations
indicate that thermal decomposition of $\rm Na_{0.7}CoO_2$ is accompanied
by a weight loss down to 96.2 wt.\% and the compound decomposes into a
sodium-rich liquid and a cobalt-rich solid phase \cite{12,13}, assuming
chemical reaction of the melt as follows, $\rm Na_xCoO_2$ $->$ CoO + Liquid
(Na-rich) + \emph{x}O$\rm _2$ - (T$>$1100 $\rm ^\circ$C) $->$ [$\rm
Na_xCoO_{1.65}$]* (*=homogeneous melt).

The dissolution of CoO occurs in the Na-rich melt and by taking up oxygen
from the environment. Thus, it results in an increase of weight because the
oxidation state of cobalt is significantly higher in the melt, i.e., $\rm
Co^{+2.7}$, according to the chemical reaction equation shown above. A
sharp weight loss down to 95.5 wt.\% occurs at $\rm T_2$=1092 $\rm ^\circ$C
and the melt starts to get homogeneous and stable. A nearly constant weight
of 95.5 - 94.9 wt.\% is obtained in the temperature range of 1120 - 1200
$\rm ^\circ$C, where the melt becomes homogeneous. A monotonic weight loss
of melt may happen under constant heating. Evidently, the melting behavior
described so far indicates that the compound melts incongruently.

\subsection{Crystal growth, morphology and composition}
When sintering $\rm Na_xCoO_2$ at high temperatures the decomposition of
$\rm Na_2O$ can cause a weight loss and lead to an inhomogeneous compound.
During growth a white powder of $\rm Na_2O$ is observed to volatilize and
deposit on the inner wall of the quartz tube. By weighing a total weight
loss $\approx$6 wt.\% is estimated. This value is in agreement with the
data shown in Fig. \ref{1}(b), neglecting the tiny loss of oxygen caused by
the change of cobalt valence state. The main loss of $\rm Na_2O$ is found
during the premelting procedure, estimated to be $\approx$5 wt.\%. This is
probably caused by incompletely reacted $\rm Na_2O$ when the mixtures are
calcined and sintered prior to premelting. An additional $\approx$1 wt.\%
loss is estimated after the crystal growth.

During growth, we observed that the molten zone is rather stable and $\rm
Na_xCoO_2$ single crystals with a sodium content of \emph{x}=0.70 - 0.90 is
easily formed. In contrast, it is hard to grow the compound with
\emph{x}=0.50 - 0.60. The analysis of EDX indicates that as-grown $\rm
Na_xCoO_2$ with \emph{x}$<$0.6 consists of multi phases like, $\rm Na_2O$,
$\rm CoO_2$ and Na-poor phases.

\begin{figure}
\includegraphics[width=0.7\linewidth]{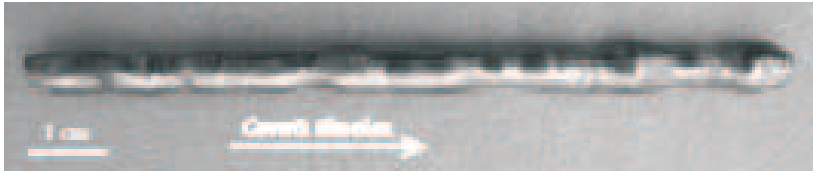}
\caption{ A typical $\rm Na_{0.7}CoO_2$ single crystal ingot obtained by
optical floating zone technique.} \label{2}
\end{figure}

\begin{figure}
\includegraphics[width=0.8\linewidth]{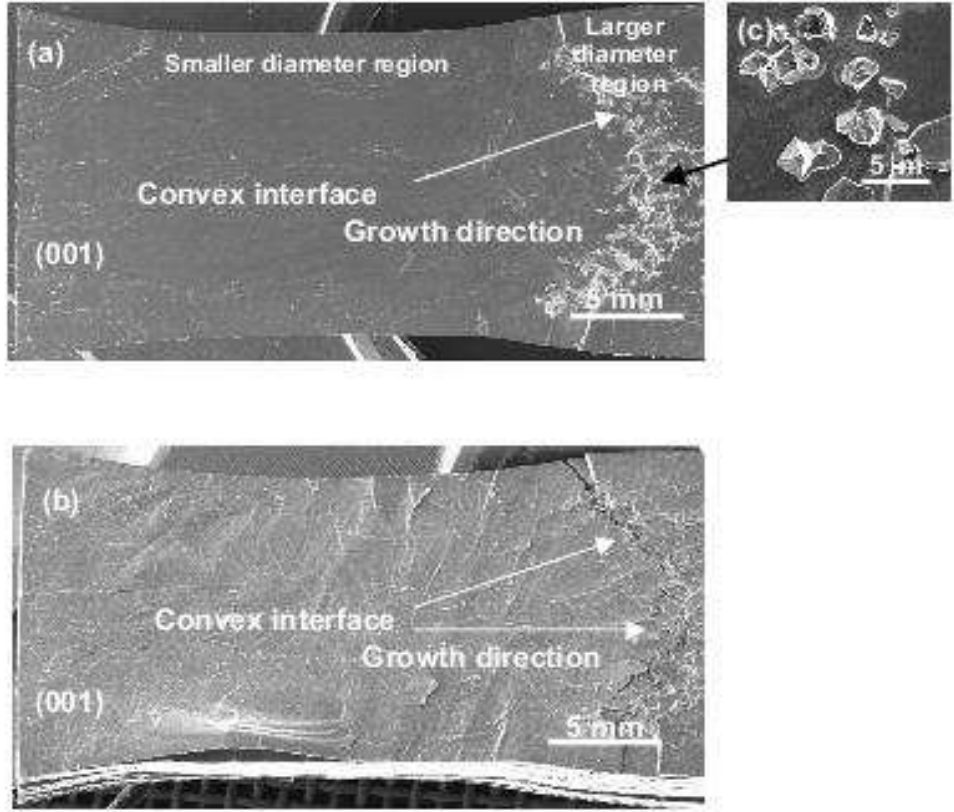}
\caption{ Two halves of an as-grown crystal ingot with the \emph{001}
surface. (a) the as-cleaved half $\rm Na_{0.7}CoO_2$ showing the $\rm
CoO_2$ inclusions gathering at the growth boundary and (b) the other half
$\rm Na_{0.3}CoO_2 yH_2O$ showing the removal of the $\rm CoO_2$ inclusions
after deintercalation followed by hydration, (c) the enlarged $\rm CoO_2$
inclusions.} \label{3}
\end{figure}

Large and high quality single crystals were obtained with a growth rate of
$\approx$2 mm/h in flowing oxygen atmosphere. Figure \ref{2} shows a
typical crystal ingot of $\rm Na_{0.7}CoO_2$. The \emph{001} face is
readily cleaved mechanically due to the layered structure of the compound.
Figure \ref{3}(a) displays one halve of the crystal platelets with the
\emph{001} face of several $\rm cm^2$ area cleaved from an ingot with a
sharp scalpel. Figure \ref{3}(b) shows the other half after water
intercalation. Crystal grains are found to grow preferably along the
\emph{a} crystallographic axis, parallel to the rod axis.

A convex growth interface is observed to be the boundary between the
regions of smaller and larger diameter of the crystal ingot, as shown in
Fig. \ref{3}. The smaller one is formed when the molten zone was at lower
temperatures and the larger one is formed at higher temperatures.
Therefore, the unequal diameter for an ingot indicates that temperature
fluctuations occurred at the molten zone during growth. According to the
temperature-composition relationship a fluctuation of heating temperature
may result in a composition change of the molten zone, leading to an
inhomogeneous compound. Figure \ref{3} (a) shows that many tiny crystal
grains of $\rm CoO_2$ gather at the boundary of the growth front. These
grains can be removed after the deintercalation treatment, as shown in
Figs. \ref{3} (b) and (c), respectively. It is important to maintain a
stable molten zone by applying a constant heating temperature for growing a
high quality single crystal.

Single crystal XRD powder diffraction of $\rm Na_{0.7}CoO_2$ gives the
following lattice parameters and cell volume of the space group P63/mmc:
\emph{a}=2.8278(4) \AA, \emph{c}= 10.9073(0) \AA, \emph{v}=75.54 \AA$\rm
^{3}$. Sintered powders show a slightly different XRD patterns with
\emph{a}=2.8230(2) \AA, \emph{c}= 10.9628(8) \AA~ \cite{1}.

For the determination of Na/Co compositions using EDX the crystal was
scanned through a segment of 3~mm along the growth direction taking the
average of four measured points in the central and edge region of the
crystal. It is found that the Na content varied with the temperature
fluctuations during growth. At the beginning of the growth the temperature
fluctuations are high and cause a high variation of the composition,
$\Delta$x$\approx$0.11, determined within 2~cm from the seeding part of the
ingot. Following the seeding part the variation of Na content is smaller,
$\Delta$x$\approx$0.06, when the molten zone is maintained in a stable
state by a constant temperature. A volatilized white $\rm Na_2O$ powder
accumulates on the inner wall of the quartz tube after the growth is
completed. The loss of Na may result to a reduction of its concentration in
the as grown crystal. The white $\rm Na_2O$ powders also form on the
surface of crystals if the samples are stored under ambient conditions.
Therefore the grown crystals must be stored in an evacuated container or a
desiccator to avoid decomposition.

\subsection{Sodium extraction and water intercalation}
The superconducting phase $\rm Na_xCoO_2 yH_2O$ (0.26$\leq$ \emph{x}$\leq$
0.42, y$\approx$1.3) is obtained by chemically extracting (deintercalation)
sodium from $\rm Na_xCoO_2$, followed by hydration. Single crystals of $\rm
Na_{0.7}CoO_2$ were placed in the oxidizing agent $\rm Br_{2}/CH_{3}CN$ for
around 100~hours, and then washed with acetonitrile. The change of sodium
content of the resulting crystals is generally proportional to the bromine
concentration in the $\rm  CH_3CN$ agent \cite{3}. A crystal of $\rm
Na_xCoO_2$ with \emph{x}=0.3 can result from a 6.6~mol $\rm Br_2/CH_3CN$
treatment only after a long extraction time of more than a week. The Na
extraction of the compound was also carried out by the electrochemical
method. The composition $\rm Na_{0.3}CoO_2$ could be achieved in aqueous
solution of NaOH using a constant current of 0.5~mA and a voltage of 1.0~V
for over 10 days. Comparing the published 1 to 5 days extraction time for
the $\rm Na_{0.7}CoO_2$ powders \cite{1,3} the crystal needs longer time to
complete the extraction, also depending on the dimension of the sample. The
slower extraction reaction in single crystals compared to a powder is
related to the higher perfection and longer period of Na-O-Na within the
NaO layers. In powder samples variations of the bonding are expected on a
nanometer scale. Before and after the extraction treatment the sodium
composition distribution across the crystals was determined by EDX.



\begin{figure}
\includegraphics[width=0.7\linewidth]{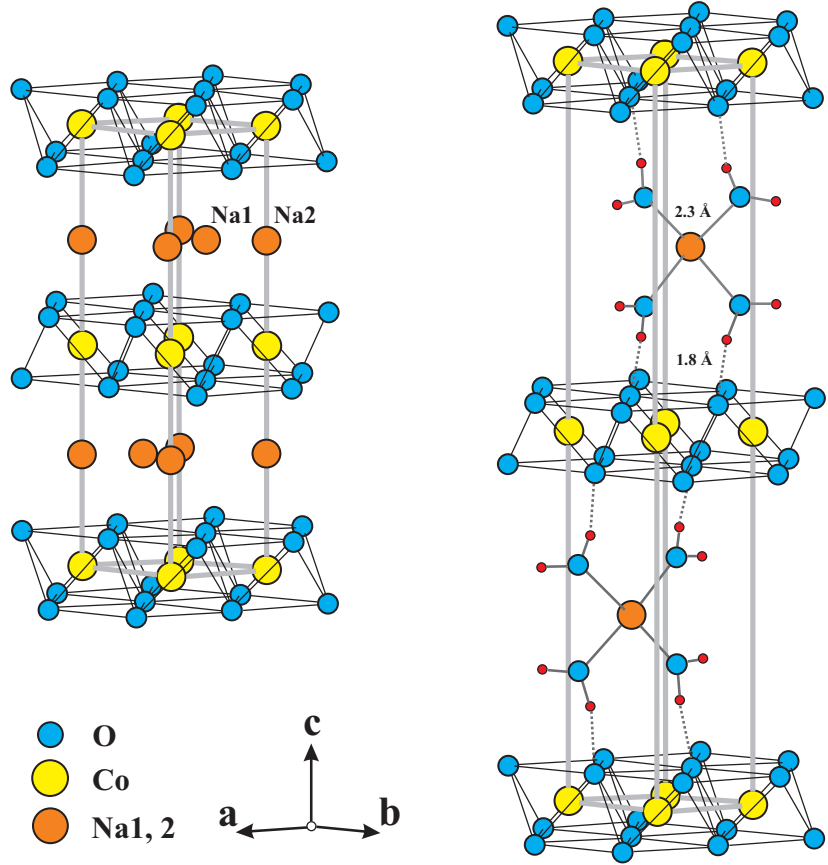}
\caption{ Schematic drawing of the layered structures \cite{14} for (a)
non-hydrated $\rm Na_{0.61}CoO_2$ and (b) fully hydrated $\rm
Na_{0.33}CoO_21.3H_2O$. } \label{5}
\end{figure}

\begin{figure}
\includegraphics[width=0.6\linewidth]{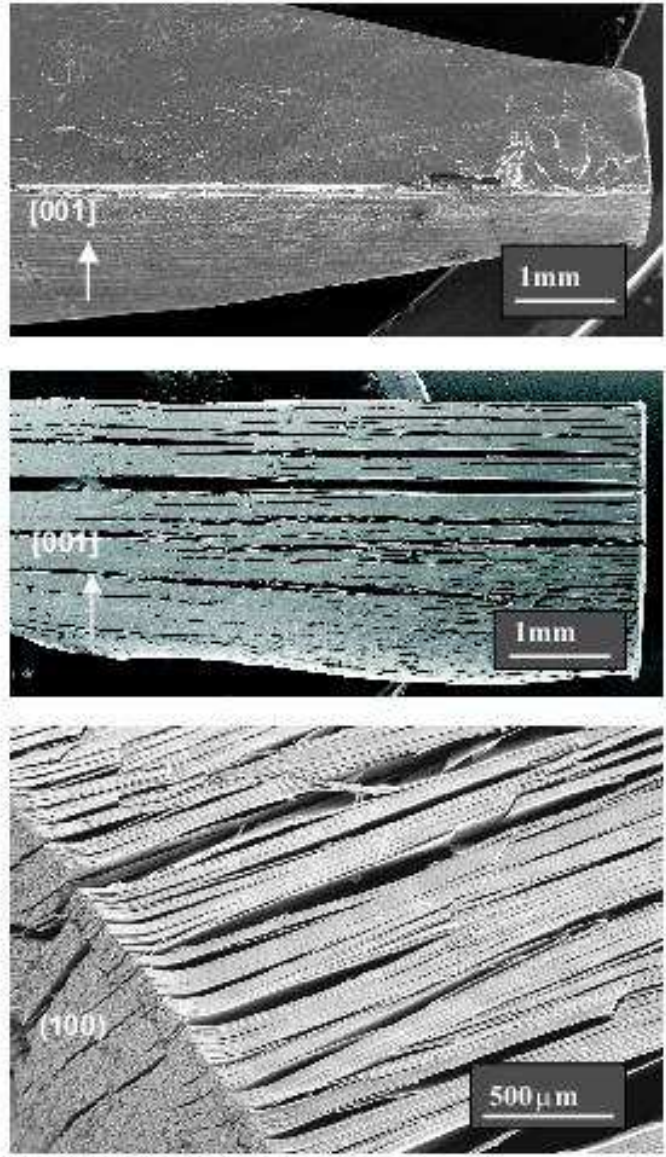}
\caption{ Three typical morphologies showing the layered structure of $\rm
Na_xCoO_2$ single crystal. (a) After cutting an as-grown $\rm
Na_{0.7}CoO_2$ single crystal, (b) the cracked layers perpendicular to the
\emph{c} axis of the $\rm Na_{0.3}CoO_{2}$ crystals after deintercalation,
(c) the "booklet"-like structure of the $\rm Na_{0.3}CoO_21.3H_2O$ after
hydrated by $\rm H_2O$.} \label{6}
\end{figure}

The superconducting phase is then obtained by immersing the deintercalated
samples for a week in $\rm H_2O/D_2O$, or sealing the samples in a water
vapor saturated container at room temperature. After hydration/deuteration
a large increase in thickness is visible to the naked eyes and the
morphology exhibits "booklet"-like layered cracks perpendicular to the
\emph{c} axis. The sodium intercalant layer expands with decreasing Na
content, because the Na removal results in Co oxidation ($\rm Co^{3+}$ ions
oxidized to smaller $\rm Co^{4+}$ ions) and thus the $\rm CoO_2$ layers are
expected to shrink \cite{1,2}. This suggests that a decreased bonding
interaction between layers with decreasing Na content may result in a
readily cleaving plane. Figures \ref{5}(a) and (b) illustrate the layered
structures of the non- and fully hydrated phases of the compound,
respectively. According to the layered structure of $\rm Na_xCo_2$
\cite{14}, 24 Co-O bonds in the layer of $\rm CoO_2$ form an octahedral
$\rm CoO_6$ with high bonding energy and therefore the layer is
structurally robust, while the bonding energy is much weaker in the NaO
layer because the $\rm Na^{+}$ mobility is high and the number of bonds
Na1-O is 6 and Na2-O only 4/3. Therefore the Na layer readily collapses to
leave the $\rm CoO_2$ layer terminated as the outer-most surface of the
crystal. Moreover, the insertion of water induces two additional
intercalated layers of $\rm H_2O$ between NaO and $\rm CoO_2$, i.e., from
Co-Na-Co to $\rm Co-H_{2}O-Na-H_{2}O-Na-Co$, which results in the expansion
of the c lattice constant from 11.2 to 19.6~\AA. The hydrogen in the new
intercalated $\rm H_2O$ layer bonds extremely weakly to the NaO and CoO
layers. Therefore this compound is exceptionally unstable with respect to a
change of thermal, mechanical or humidity environment. Under ambient,
oxidizing and humid conditions the crystal exhibits three typical
morphologies, as shown in Figs. \ref{6}(a), (b) and (c), respectively.

\begin{figure}
\includegraphics[width=0.7\linewidth]{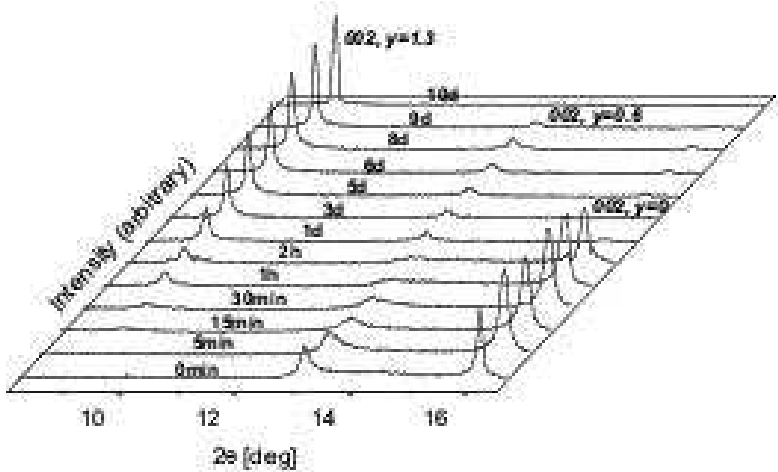}
\caption{ Single crystal X-ray diffraction patterns for $\rm Na_{0.3}CoO_2
yH_2O$ measured by immersing the sample in D.I. water. The initial crystal
contained the phases of \emph{y}=0.6 and 0. } \label{7}
\end{figure}

\subsection{Behavior of the crystals upon hydration and dehydration }
To study the phase formation and problems with rehydration we have
performed detailed x-ray diffraction experiments as function of time.
Single crystals of non-hydrated $\rm Na_{0.3}CoO_2$ have been partially
hydrated in humid air to form $\rm Na_{0.3}CoO_20.6H_2O$ under ambient
conditions. These crystals were immersed in water with the \emph{c} surface
exposed to air. Each measurement of the \emph{002} reflections was carried
out for a short time of $\approx$3 minutes to avoid that the x-ray emits
off the sample surface due to the rapid expansion of crystal thickness
during the hydrating process. We show in Fig. \ref{7} that the increase of
intensity of the \emph{002} reflection (corresponding to \emph{y}=1.3) is
accompanied by a decrease of intensity of the other two \emph{002}
reflections (corresponding to \emph{y}=0.6 and 0, respectively). This is
indicative of the increase of the volume of hydrated phase \emph{y}=1.3
while decreasing the non-hydrated and partial hydrate phases, \emph{y}=0
and 0.6. After a one day hydration the sample becomes completely "wet".
This is indicated by the \emph{002} reflections for the \emph{y}=1.3 and
0.6 phases and the disappearance of the \emph{002} corresponding to the
non-hydrated phase \emph{y}=0. Note that the \emph{002} reflection of the
partial hydrate with \emph{y}=0.6 still exists until a further hydration up
to 10 days. In the meantime the XRD pattern shows a continuously increasing
volume fraction of \emph{y}=1.3 phase reaching approximately 100\%. This
suggests that the initial hydration process takes place with two-water
molecules per unit cell (corresponding to \emph{y}=0.6) inserted into the
Na plane, and followed by a group of four (corresponding to \emph{y}=1.3)
located in vacancies of the extracted Na sites.

The hydrated volume of a crystal is a function of hydration time since the
water diffuses gradually in the \emph{ab} plane whereas the \emph{c} face
is robust and no diffusion path exists along the \emph{c} axis. A fully
hydrated phase \emph{y}=1.3, exhibiting superconductivity, is achieved
after 10 days for a crystal of 2x2 $\rm mm^{2}$ in the \emph{a}/\emph{b}
direction, concomitant with the disappearance of the partial hydrates of
the other two \emph{002} reflections, as shown in Fig.~\ref{7}. This
indicates that Na-vacant sites can be completely occupied by water
molecules when a longer diffusion time is allowed for creating a
water-saturated environment during the hydration process.

\begin{figure}
\includegraphics[width=0.85\linewidth]{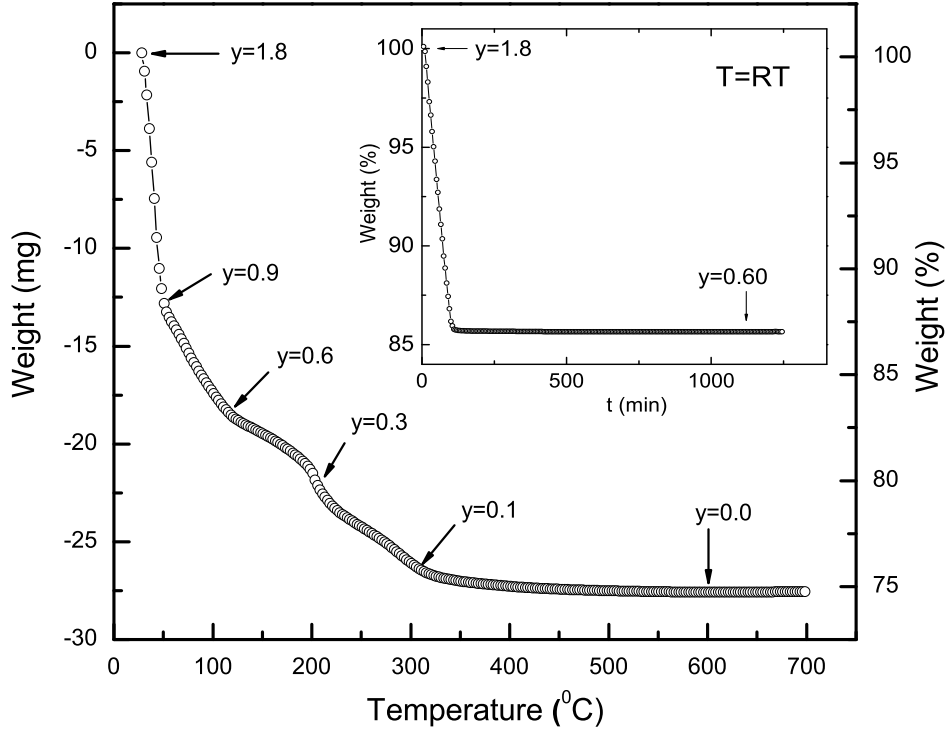}
\caption{ Thermogravimetric analysis of an over hydrated $\rm
Na_{0.3}CoO_21.8H_2O$ single crystal heated at 0.3 K/min in flowing oxygen.
Inset: time dependence of the weight loss for the same specimen at room
temperature in flowing oxygen. Initial mass of the sample was 112.965~mg.
The loss was approximately 19 mg after heating and rehydrating.} \label{8}
\end{figure}

A thermogravimetric analysis can be used to monitor the loss of water and
stability of the hydrated phases as function of temperature and time.
Figure \ref{8} shows the temperature dependence of water loss for an over
hydrated sample (\emph{y}=1.8) under a slow heating rate of 0.3 $\rm
^\circ$C /min in flowing oxygen. The fastest loss of water occurs between
20 and 50 $\rm ^\circ$C and is then again increasing upon further heating.
The sample weight is characterized by several relatively broad plateaus in
weight that are attributed to phases with \emph{y}=1.8, 0.9, 0.6, 0.3 and
0.1 for temperatures near 28, 50, 100, 200 and 300 $\rm ^\circ$C,
respectively. These data suggest the existence of the following majority
phases at the respective temperatures: $\rm Na_{0.3}CoO_21.8H_2O$
(\emph{c}=19.75 \AA), $\rm Na_{0.3}CoO_20.9H_2O$ (\emph{c}=13.85 \AA), $\rm
Na_{0.3}CoO_20.6H_2O$ (\emph{c}=13.82 \AA) and $\rm Na_{0.3}CoO_20.3H_2O$
(\emph{c}=12.49\AA) and $\rm Na_{0.3}CoO_20.1H_2O$ (\emph{c}=11.20\AA). A
change of $\Delta$y=0.33 corresponds to the removal of one water molecule
per unit cell. A complete removal of water (\emph{y}=0) from the compound
takes place at temperatures over 600 $\rm ^\circ$C.

Analysis of DTA-TG confirms that the fully hydrated superconducting phase
(\emph{y}=1.3) is rather unstable between 20 and 50 $\rm ^\circ$C due to
the rapid loss of water. However, the superconducting phase can be reformed
easily by rehydration. The inset of Fig. \ref{8} shows the time dependence
of the loss of water at room temperature. A rapid loss of water from
\emph{y}=1.8 to 0.6 occurs in one and a half hours and then no further
weight loss is observed during one day. XRD reveals that the flat plateau
corresponding to a semi hydrated $\rm Na_{0.3}CoO_20.6H_2O$ with
\emph{y}=0.6 appears to be a metastable phase under ambient conditions.
This observation is in conflict with that for polycrystalline powders
\cite{2}.

The fact of the matter is that a single crystal is only one domain
containing crystal-water while powder samples consist of countless tiny
grains with additional intergrain water. Furthermore, the loss or
absorption of water in a large single crystal is slower and more
well-defined compared to powders due to the diffusion path along the
Na-layer.


\subsection{Mixture of cells}





\begin{figure}[tbp]
\includegraphics[width=0.85\linewidth]{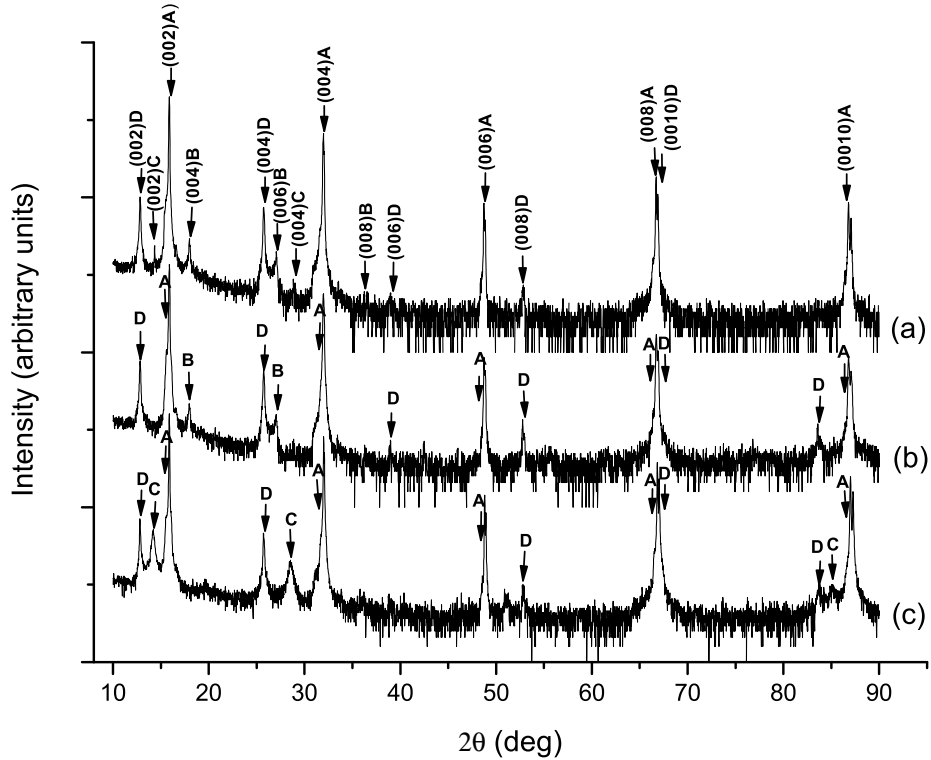}
\caption{ The \emph{00l} reflections showing the mixture of cells and the
lattice constants for the various hydrates of $\rm Na_{0.3}CoO_2 yH_2O$
under different ambient conditions. (a) From humid to dry air for 2 days,
(b) from humid to dry air for 5 days, (c) from dry to humid air for 5 days.
Symbol \textbf{A} represents the phase of \emph{y}=0 for \emph{a}=2.82 \AA~
and \emph{c}=11.195 \AA, \textbf{C}: \emph{y}=0.3 for \emph{a}=2.82 \AA~
and \emph{c}=12.478 \AA, \textbf{D}: \emph{y}=0.6 for \emph{a}=2.82 \AA~
and \emph{c}=13.815 \AA, \textbf{B}: \emph{y}=1.3 for a=2.82 \AA~ and
\emph{c}=19.710 \AA.  } \label{10}
\end{figure}

As shown in the TGA data, a fully hydrated sample can release water under
ambient conditions to form partial hydrates. On the other hand, moisture
from air can also be absorbed by a non-hydrated sample to form hydrated
phases. In order to quantitatively identify the hydrated phases in the
crystal, the fully hydrated or non-hydrated samples were placed in a humid
or dry atmosphere. They were then investigated by XRD and the results show
a clear shift in the positions of the \emph{00l} reflections for both
samples, yielding several variations in the \emph{c} axes of the unit
cells, as shown in Figs.~\ref{10}(a-c). The fully hydrated phase of $\rm
Na_{0.3}CoO_21.3H_2O$ is exposed to dry air at ambient conditions for 2 and
5 days, resulting in the formation of four and three mixed phases, as shown
in Figs.~\ref{10}(a) and (b), respectively. Exposing a non-hydrated $\rm
Na_{0.3}CoO_2$ to humid air for 5 days results in two additional phases of
$\rm Na_{0.3}CoO_20.3H_2O$ and $\rm Na_{0.3}CoO_20.6H_2O$, as shown in
Fig.~\ref{10}(c). Note that the phase with \emph{y}=0.6 is always observed
in the crystals under ambient conditions. Nevertheless, to prevent a fully
hydrated phase from losing water we suggest that the samples be stored in a
saturated humid atmosphere. Non-hydrated samples can be stored in an
evacuated container to prevent decomposition from absorbing water in air.

\begin{figure}[tbp]
\includegraphics[width=0.7\linewidth]{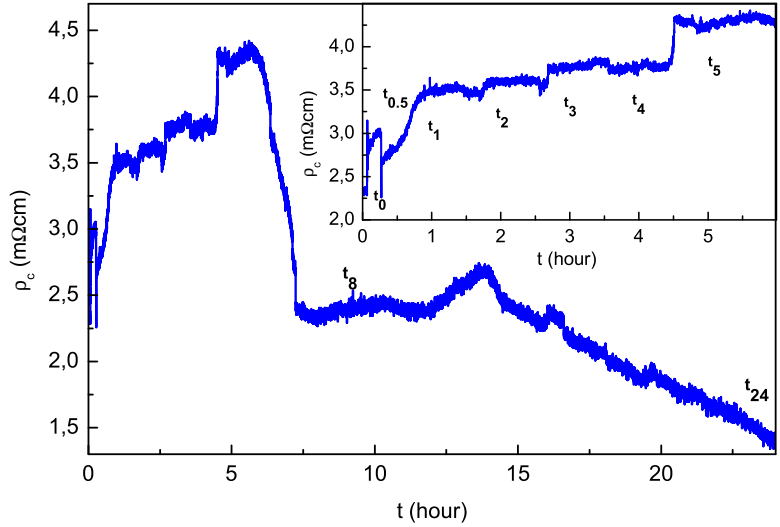}
\caption{ Time dependence of the out-of-plane resistivity $\rm \rho_c$ of a
$\rm Na_{0.3}CoO_2 yH_2O$ single crystal during a hydration process. The
number given as footnote at the plateaus is hydration time in hours.}
\label{11}
\end{figure}

\subsection{Electrical properties}
We have performed transport experiments of the out-of-plane resistivity
$\rm \rho_c$ as function of time to further characterize the hydration
dynamics. The initially non-hydrated single crystal $\rm Na_{0.3}CoO_2$ was
immersed in D.I. water throughout the whole experiment. In contrast to
measurements of powder samples in humid air, this configuration reduces the
influence of internal and external surfaces and allows a full and
continuous hydration. Furthermore, the out-of-plane resistivity $\rm
\rho_c$ measured in four-point geometry should be less affected by the
intrinsic Na ion conductivity.

In Fig. \ref{11} and its inset plateaus and steps of the resistivity are
given with increasing hydration time followed by a continuous and then more
gradual decrease. Noteworthy is the similar resistivity reached after 8
hours compared to the initial state. This implies that, although
irreversible by nature, the hydration process has also reversible stages.
During hydration the water molecules enter the compound to intercalate
between the $\rm CoO_2$ and Na layers. The $\rm CoO_2$ layers are separated
by a trilayer of $\rm H_2O$/Na/$\rm H_2O$. The dominant effect is to expand
the c-axis lattice parameters, from about 11.2 \AA~ without water
(non-superconducting phase $\rm Na_{0.3}CoO_2yH_2O$, \emph{y}=0) to 12.5
\AA~ (semi hydrated phase, \emph{y}=0.6) and then to 19.5 \AA~ when the
system becomes superconducting (\emph{y}=1.3) (see Fig. \ref{7}). This is
related to an enlarged interlayer distance from 5.6 \AA~ to 6.9 \AA~ and
then to 9.8 \AA. The stepwise increase from $\rm t_0$ to $\rm t_5$ occurs
at a constant chemical doping level and suggests that the increase of
spacing between the $\rm CoO_2$ layers leads to the resistivity increase.
We propose that the rapid steps followed by plateaus map the dynamics of
domains of hydration that are pinned at the random potential of the local
defect landscape. The more gradually decrease of resistivity for
t$>$12~hours must therefore be related to a more homogeneous state of the
system.

There is an interesting correlation between the $\rm CoO_2$ layer and the
superconducting transition temperature, i.e., the Co-O distance decreases
with increasing $\rm T_c$ ~\cite{14}. The layer thickness of $\rm CoO_2$
changes from 1.93 \AA~ for non-superconducting phase to 1.84 \AA~ for
superconducting phase \cite{15}. It is clear that the layer thickness of
$\rm CoO_2$ shrinks after reaching full hydration or doping to \emph{y}=1.3
and a change in the oxidation state of Co occurs, because the oxidation
state directly corresponds to the carrier density in the $\rm CoO_2$ layers
and some electron transfer off the cobalt \cite{1}. Recently, also a
relation between the \emph{c} axis parameter and the superconducting
transition temperature has been discussed \cite{milne04}. Nevertheless, the
change of the room temperature resistivity during the hydration process is
not only influenced by the expansion of spacing in the system but also by a
redistribution of the charge in $\rm CoO_2$ layer.

\subsection{Superconductivity}

\begin{figure}[tbp]
\includegraphics[width=0.8\linewidth]{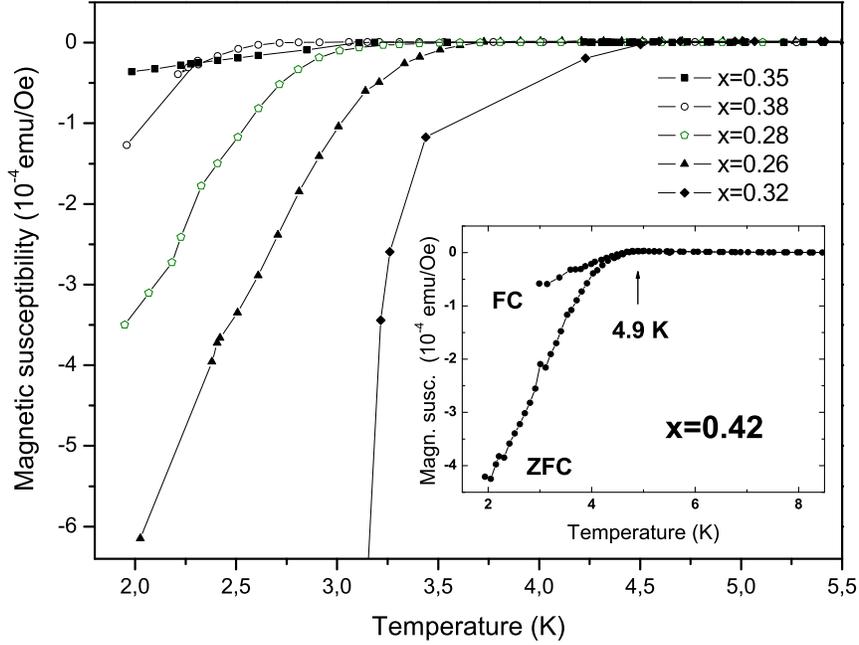}
\caption{ Zero field cooling (ZFC) magnetic characterization of the $\rm
Na_{x}CoO_21.3H_2O$ single crystals showing the onset $\rm T_c$(x). Inset:
optimum $\rm T_c \approx$4.9~K with x=0.42 as ZFC and FC measurement.  }
\label{12}
\end{figure}

The superconducting properties of the fully hydrated single crystals were
characterized by a.c. susceptibility using a SQUID magnetometer. Zero field
cooling measurements with 10~Oe applied field are presented in
Fig.~\ref{12}. Superconducting transition temperatures $\rm T_c$=2.8 -
4.9~K with \emph{x}$\approx$0.28 - 0.42 are observed, respectively. Some of
our samples showed strong diamagnetic a.c. signals and a superconducting
volume fraction as high as 20\%. The transition width for each sample is
different, which is very likely caused by an inhomogeneous sodium
concentration and partial hydration that affect the superconducting state
of $\rm Na_{x}CoO_21.3H_2O$. The $\rm T_c$ values vary with Na content,
showing the highest $\rm T_c$ at 4.9K with \emph{x}=0.42 given in the inset
of Fig.~\ref{12}. As shown in the inset the onset of the superconducting
transition is identical with the divergence of FC and ZFC data.

\begin{figure}[tbp]
\includegraphics[width=0.7\linewidth]{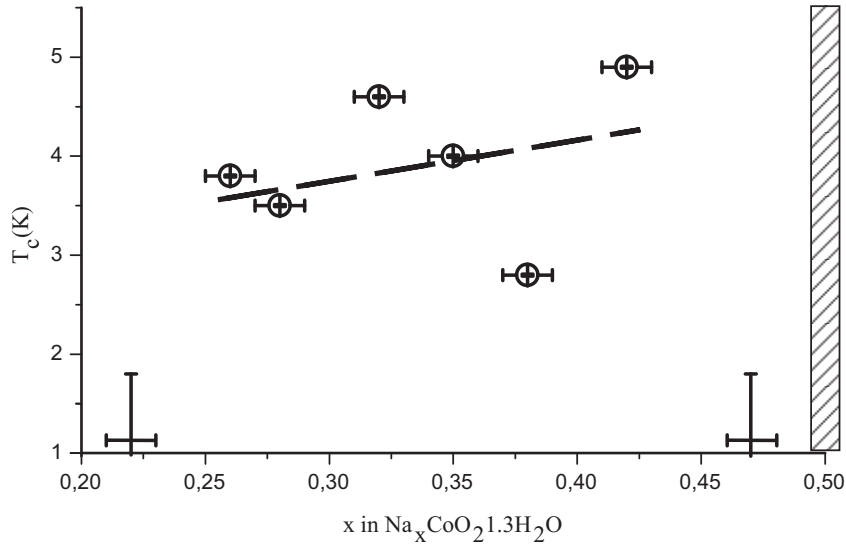}
\caption{ Superconducting phase diagram as a function of Na content. The
dashed line is a guide to the eye. The dashed bar marks the metal-insulator
transition observed in non-hydrated $\rm Na_{0.5}CoO_2$.  } \label{13}
\end{figure}

Figure \ref{13} is the plot of $\rm T_c$ as function of sodium
concentration of $\rm Na_{x}CoO_21.3H_2O$. For samples with
\emph{x}$\leq$0.22 and \emph{x}$\geq$0.47 there is no superconducting
transition detectable. These data agree to some extend with a recent study
that shows a constant $\rm T_c$ for Na contents up to 0.37 \cite{milne04}.
However, they conflict with the superconducting "dome" of $\rm T_c$(x)
demonstrated for powder samples \cite{3}. This latter dependence motivated
the proposal of intrinsic critical concentrations
\emph{x}$\rm_{cr1}$$\approx$1/4 and $x\rm _{cr2}$$\approx$1/3 related to
charge ordering instabilities of the $\rm Co^{3+}/Co^{4+}$
~\cite{baskaran04}. In the present single crystal study the upper limit is
shifted to much higher concentrations, \emph{x}$\rm _{cr2}$ $\approx$ 0.45.
This implies, that a charge ordering at \emph{x}$\rm _{cr2}$$\approx$1/3 is
not relevant to suppress $\rm T_c$.

However, the superconducting phase extends now close to the phase line
where non-hydrated $\rm Na_{0.5}CoO_2$ shows a metal-insulator transition
\cite{huang04}. This is evidence for an importance of electronic
correlations at higher \emph{x} in the hydrated system. It is interesting
to note that the N\'{e}el transition $\rm T_{N}$$\approx$20K is also only
observed for higher Na concentrations \emph{x}$>$0.75 in non-hydrated
crystals, and that this ordering temperature does not change appreciably
with \emph{x}. Bulk antiferromagnetic order has been proven using myon spin
rotation and other thermodynamic experimental technique \cite{6}. The exact
nature of the ordering, however, is still needed to further investigate.

Evidence for superlattice formation and electronic instabilities either due
to Na or $\rm Co^{3+}/Co^{4+}$ charge ordering are recently accumulating
for non-hydrated $\rm Na_{x}CoO_2$ with \emph{x}=0.5, but also to some
extend for other stoichiometries \cite{zandbergen04,gavilano04}. Hydration
contributes to the charge redistribution due to its effect on the $\rm
CoO_2$ layer thickness and the formation of $\rm Na(H_2O)_4$ clusters that
shield disorder of the partially occupied Na sites. The above mentioned
instabilities should influence the superconducting state as they modulate
the electronic density of states and change nesting properties of the Fermi
surface \cite{tanaka03,johannes04}. Further studies are needed to
investigate a possible interrelation of Na content, hydration and
structural details on well-characterized samples with highest $\rm T_c$.

\section{Conclusion}
Using TG-DTA we found that $\rm Na_xCoO_2$ is an incongruently melting
compound, and single crystals can be grown using the optical floating zone
technique. Both non-hydrated and fully hydrated crystals are exceptionally
sensitive under ambient conditions. Study of XRD and TG indicates that the
semi-hydrated phase $\rm Na_{x}CoO_20.6H_2O$ is more stable than the other
compositions under ambient conditions. The XRD patterns show the presence
of cells corresponding to the coexistence of several different hydrated
phases in rehydrated samples. The superconducting transition temperature of
$\rm Na_{x}CoO_21.3H_2O$ is only weakly influenced by the Na content.
Highest $\rm T_c \approx$4.9~K is found for x$\approx$0.42, i.e. in the
proximity of a metal-insulator transition observed in the non-hydrated
phase with \emph{x}=0.5. The hydrated compound is strongly unstable
concerning its chemical, structural and electrical properties. We therefore
propose to carefully reinvestigate the recently claimed evidence for
unconventional superconductivity on single crystals.

{\bf Acknowledgements:} We thank E. Br\"{u}cher, G. G\"{o}tz, E. Peters,
and E. Winckler for technical support.


\begin{thebibliography}{10}

\bibitem{1} K. Takada, H. Sakurai, E. Takayama-Muromachi,
F. Izumi, R. Dilanian, and T. Sasaki, Nature, \textbf{422}, 53 (2003).

\bibitem{2} M.L. Foo, R.E. Schaak, V.L. Miller, T. Klimczuk,
N.S. Rogado, Yayu Wang, G.C Lau, C. Craley, H.W. Zandbergen, N.P. Ong, and
R.J. Cava, Solid State Comm. \textbf{127}, 33(2003).

\bibitem{3}  R.E. Schaak, T. Klimczuk, M.L. Foo, and R.J. Cava,
Nature, Vol. \textbf{424}, 527 (2003).

\bibitem{4} T. Fujimoto, G.Q. Zheng, Y. Kitaoka, R.L. Meng, J. Cmaidalka, and C.W Chu,
cond-mat/0307127 v1 6Jun2003.

\bibitem{5} R. Jin, B.C. Sales, P. Khalifah, and D. Mandrus,
cond-mat/0306066 v1 3Jun2003.

\bibitem{6} S. Bayrakci, C. Bernhard, D.P. Chen, B. Keimer,
P. Lemmens, C.T. Lin, C. Niedermayer, and J. Strempfer,
cond-mat/0312376,
to be published in Phys. Rev. B. (2004) and references within.

\bibitem{7} P. Lemmens, V. Gnezdilov, N.N. Kovaleva, K.Y. Choi,
H. Sakurai, E. Takayama-Muromachi, K. Takada, T. Sasaki, D.P. Chen, F.C.
Chou, C.T. Lin, B. Keimer, Journ. Phys.: Cond. Matt. \textbf{16}, S857
(2004) and references therein.

\bibitem{tanaka03} A. Tanaka and X. Hu,
Phys. Rev. Lett \textbf{91}, 257006 (2003).

\bibitem{johannes04}M.D. Johannes, I.I. Mazin,
D.J. Singh, D.A. Papaconstantopolus, cond-mat/0403135, (2004).

\bibitem{huang04}   Q. Huang, M.L. Foo, J.W. Lynn,
H.W. Zandbergen, G. Lawes, Yayu Wang, B. H. Toby, A.P. Ramirez, N.P. Ong,
and R.J. Cava, cond-mat/0402255, (2004).

\bibitem{8} F. Rivadulla, J.-S. Zhou, J. B. Goodenough, cond-mat/0304455.

\bibitem{9} T. Motohashi, E. Naujalis, R. Ueda, K. Isawa,
M. Karppinen, and H. Yamauchi, Appl. Phys. Lett. \textbf{79}, 1480 (2001).

\bibitem{10}  K. Fujita, T. Mochida and K. Nakamura,
Jpn. J. Appl. Phys. \textbf{40} 4644 (2001).

\bibitem{11} F.C. Chou, J.H. Cho, P.A. Lee, E.T. Abel,
K. Matan, and Y.S. Lee, cond-mat/030606659,
to be published in Phys. Rev. \textbf{B} (2004).

\bibitem{12}  M. Chen, B. Hallstedt and L.J. Gauckler,
J. phase Equil. 2003(\textbf{24}) p.212-227.

\bibitem{13} Wriedt, Bull. Alloy Phase Diagrams 1987(8) p. 234-245.

\bibitem{14} J.D. Jorgensen, M. Avdeev, D.G. Hinks, J.C. Burley,
S. Short, Phys. Rev. \textbf{B 68}, 214517 (2003).

\bibitem{15} J.W. Lynn, Q. Huang, C.M. Brown, V.L. Miller,
M.L. Foo, R.E. Schaak, C.Y. Jones, E.A. Mackey, and R.J. Cava,
Phys. Rev. \textbf{B 68}, 214516 (2003).

\bibitem{milne04}  C. J. Milne, D. N. Argyriou, A. Chemseddine,
N. Aliouane, J. Veira, and D. Alber, cond-mat/0401273.

\bibitem{baskaran04}  G. Baskaran, cond-mat/0306569,
Proceedings of M2S-Rio 2003.

\bibitem{zandbergen04}H.W. Zandbergen, M.L. Foo, Q. Xu, V. Kumar and R.J.
Cava, cond-mat/0403206 (2004.

\bibitem{gavilano04} J.L. Gavilano, D. Rau, B. Pedrini, J. Hinderer,
H.R. Ott, S.M. Kazakov and J. Karpinski, cond-mat/0308383, (2003).




\end{thebibliography}
\end{document}